\documentclass[letterpaper, 10 pt, conference]{ieeeconf}  

\IEEEoverridecommandlockouts                              
\usepackage{booktabs} 
\usepackage{graphicx}
\usepackage{color}
\usepackage{amssymb}
\usepackage{threeparttable}
\usepackage{amsmath}
\usepackage{soul}
\usepackage{caption}
\usepackage{subcaption}
\usepackage{float}
\usepackage{booktabs}
\usepackage{stfloats}

\title{\LARGE \bf
Text Extraction and Retrieval from Smartphone Screenshots: Building a Repository for Life in Media
}

\author{
Agnese Chiatti\textsuperscript{\textdagger}
Mu Jung Cho**
Anupriya Gagneja**
Xiao Yang*
Miriam Brinberg*\\
Katie Roehrick**
Sagnik Ray Choudhury\textsuperscript{\textdagger} 
Nilam Ram*
Byron Reeves**
C. Lee Giles\textsuperscript{\textdagger} 
}

\begin{document}

\maketitle
\thispagestyle{empty}
\pagestyle{empty}

\noindent\textsuperscript{\textdagger}Information Sciences and Technology, Pennsylvania State University, 
\{azc76,szr163,giles\}@ist.psu.edu\\
\noindent*Human Development and Family ~ Studies, Pennsylvania State University, \{xfy5031,mjb6504,nur5\}@psu.edu\\
\noindent**\{Department of Communication, Department of Computer Science\}, Stanford University,\\
\noindent\{mujung.cho,anupriya,kroehr,reeves\}@stanford.edu\\

\begin{abstract}
Daily engagement in life experiences is increasingly interwoven with mobile device use. 
Screen capture at the scale of seconds is being used in behavioral studies and to implement "just-in-time" health interventions. The increasing psychological breadth of digital information will continue to make the actual screens that people view a preferred if not required source of data about life experiences.
Effective and efficient Information Extraction and Retrieval from digital screenshots is a crucial prerequisite to successful use of screen data. In this paper, we present the experimental workflow we exploited to: (i) pre-process a unique collection of screen captures, (ii) extract unstructured text embedded in the images, (iii) organize image text and metadata based on a structured schema, (iv) index the resulting document collection, and (v) allow for Image Retrieval through a dedicated vertical search engine application. The adopted procedure integrates different open source libraries for traditional image processing, Optical Character Recognition (OCR), and Image Retrieval. Our aim is to assess whether and how state-of-the-art methodologies can be applied to this novel data set. We show how combining OpenCV-based pre-processing modules with a Long short-term memory (LSTM) based release of Tesseract OCR, without \textit{ad hoc} training, led to a 74\% character-level accuracy of the extracted text. Further, we used the processed repository as baseline for a dedicated Image Retrieval system, for the immediate use and application for behavioral and prevention scientists. \\
We discuss issues of Text Information Extraction and Retrieval that are particular to the screenshot image case and suggest important future work.

\end{abstract}

\section{INTRODUCTION}
Individuals increasingly record, share, and search through their daily activities and behaviors through digital means. Given the breadth of current technologies, this "life in media" (i.e., daily experiences captured by technology) is fragmented across multiple devices and applications. Fast-paced switching - as quickly as every 19 seconds on laptop computers \cite{Yeykelis2014} - often occurs between either loosely-related or radically different content. However, holistic examination of individuals' mediatized lives remains largely unexplored. In this scenario, digital screenshots of individuals' devices represent an ideal target to (i) ensure in situ collection of diverse data that are normally tackled within separate domains (e.g., examining use of all social media, rather than only a particular site), (ii) investigate the interdependencies and fragmentation of activities over time, forming a valuable data set for behavioral scientists and intervention implementation, and (iii) more broadly, test theory, specifically with respect to digital life (e.g., social interactions, learning, news consumption).\\
Frequent screen capture as a data collection method has the potential for providing more fine-grained, longitudinal data than traditional URL, query logging and app usage tracking methods. 
The extraction, representation and effective retrieval of screenshot textual contents are then crucial steps eliciting the application of these analyses to a number of use cases \cite{reeves2017screenomics}, including: medical diagnosis and early detection of disease, behavioral studies interrogating on the nature and shape of human development in the digital era, models of task switching and its implications for attention and memory, mining of political attitudes and voting across social media and news media, assessments of the role of fake news in democratic settings and so forth.\\ 
As discussed in \cite{chiatti2017kcap}, screenshots provide a unique combination of graphic and scene text, motivating the evaluation of state-of-the-art OCR methods on this particular data set. Further, Information Extraction from digital screenshots raises the need for developing a general purpose framework that handles a variety of fonts and layouts, disambiguates icons and graphical contents from purely textual segments, while dealing with text embedded in advertisements, logos, or video frames. \\
In order to move from miscellaneous information fragments to a  more structured and interactive repository for further Knowledge Discovery, it is important to consider the organization and accessibility of the extracted data. We particularly focus on the retrieval of screenshot images based on their textual content and metadata, by describing the search engine architecture developed to this aim.\\
This paper presents a complete workflow (from Image Pre-processing to Image Retrieval) integrating existing open-source methods and tools. The reviewed literature and related background concepts are presented in Section 2. Section 3 illustrates the design and characteristics of the data set at hand, while also describing the Image Processing and Retrieval steps. Results from the evaluation of the recognized raw text are discussed in Section 4. We conclude with future research directions in Section 5.

\section{BACKGROUND}
Text in imagery is typically characterized as either machine-printed (\textit{graphic}) or \textit{scene} text, i.e. captured on objects and native scenes \cite{survey2015ocr}.
The overarching goal of Text Information Extraction \cite{survey2015ocr}, in both cases, is to first assess the presence of any textual contents, to localize them, and to ultimately recognize the string counterpart of each provided glyph. 
Even though Text Detection and Optical Character Recognition (OCR) have reached optimal performance on scanned documents (with recognition rates exceeding 99\%), the processing of more complex or degraded images is still gathering research interest \cite{survey2015ocr}, particularly in the context of natural scene images, ancient manuscripts, and handwritten pieces \cite{wang2012end, huang2016detecting,wang2010word}, where accuracy tends to drop around 60\%.
In the case of screenshots, some traditional issues are mitigated, e.g., diverse text orientation and uneven illumination, due to the high occurrence of "graphic text" over "scene text". However, other challenges apply, such as co-occurrence of icons and text and the variability in fonts and layouts \cite{survey2015ocr}. Further, screenshots represent a hybrid case study, mixing graphic and scene test in varying proportions over time, hence motivating the evaluation of existing techniques on a novel collection.\\
Ultimately, the extracted text and the associated metadata constitute the basis to represent and retrieve information within the discussed corpus. 
Image Retrieval can generally be based on the visual elements embedded in the image (i.e., Content-based Image Retrieval or CBIR) or on its textual metadata, e.g. tags, surrounding captions or headings. This latter branch of Information Retrieval is also known as Concept-based Image Retrieval and, when built over categorical labels, traditionally requires significant manual annotation effort, as opposed to the increased computation complexity implied by CBIR.
Multimedia content recognition and indexing is being already applied: for biomedical imagery cataloging and to assist future diagnosis based on past clinical data \cite{faria2015content,Jung:2014:SMO:2554850.2555110}; for word matching over scanned and printed document images \cite{Meshesha2008}; for similarity search over large video collections \cite{daSilva:2016:VSS:2851613.2851876}. \\
The work presented in \cite{yeh2009sikuli} shares some similarities with our current pipeline, however Yeh et al. focus on GUI sub-elements of screenshots, from a user interaction standpoint and without evaluating the OCR robustness in the case of screenshots. Recent work \cite{umemoto2017search} has focused on extracting news articles metadata from smartphone screenshots and exploit those as search parameters, to return the document full text for users to consume at a later stage. To our knowledge, none of the surveyed applications has exploited a combination of text extraction and indexing to retrieve smartphone screenshots of heterogeneous source, offering a diverse range of textual contents (e.g. social media posts, news, articles, health-related threads, text messages, video captions and so forth), and, thus, enabling countless free-text search combinations, over the archived media behaviors.

\section{DATASET AND PROCESS WORKFLOW}

\subsection{Dataset}
The procedures described in this paper were applied to a set of over one million images collected from 54 individuals participating in an on-going study of media behavior. Software was installed on participants' Android smartphones and/or personal laptops to monitor media exposure and use during a 10 day to 1 month period. Screenshots were taken every 5 seconds, encrypted, bundled, and transmitted to secure, cloud-based research servers. The interval between screenshots, and resulting volume, variety, and velocity of images, ensures sufficient granularity for time-series and mathematical modeling of these data \cite{Ram2014isahib, Yeykelis2014}. Further, device use has consolidated towards smartphones over time and offers access to a significant portion of daily activities, motivating our focus on smartphone screenshots.  
The data are unique and, after information extraction, have high-value potential for many research- and consumer-oriented applications. 

For purposes of the current paper, we selected a subset of screenshots for study, testing, and refinement of text extraction procedures. Specifically, we partitioned a random subsample of 17 participants and applied reservoir sampling to randomly select 150 images from each day, thus accommodating the fact that the size of each daily collection of screenshots is not known a priori. The analysis set here consists of 13,172 smartphone screenshots representative of a typical user's behavior. 

\subsection{Image Pre-processing}
\begin{figure*}[t]
\caption{\label{fig:framework} Overall Architecture for Smartphone Screenshot Processing, Indexing and Retrieval.}
\includegraphics[width=\textwidth]{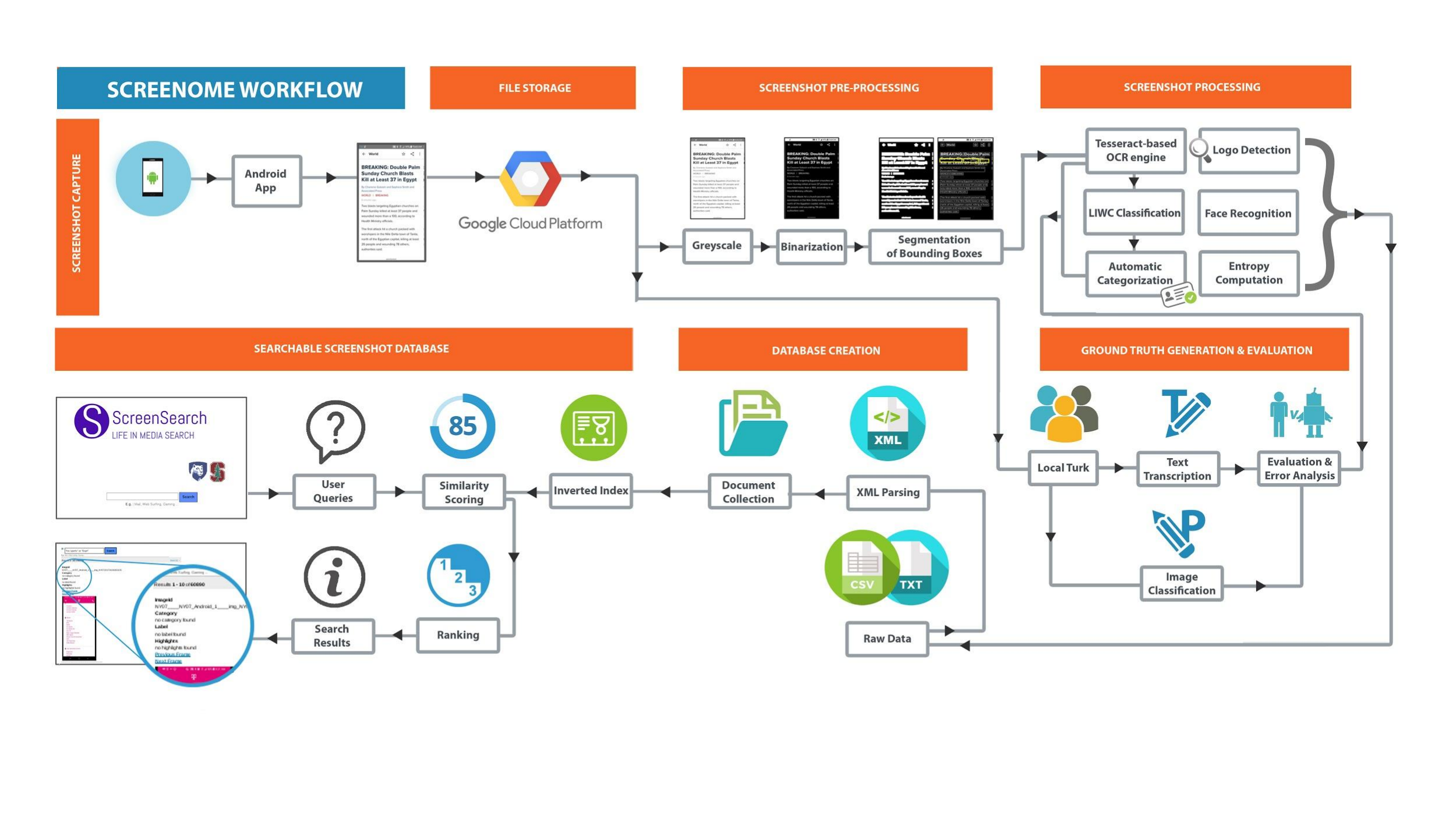}
\end{figure*}

To enhance the quality of extracted text, we setup a procedure to process the raw screenshots and feed the OCR engine with graphic segments where the textual content could be more easily distinguished from the background. This data preparation routine was built on top of the OpenCV library for Image Processing \cite{opencv_library}. The overall workflow is depicted in Figure \ref{fig:framework}, with pre-processing consisting of conversion to gray scale, binarization, and segmentation respectively.

\subsubsection{Conversion to grayscale} Conversion of the images from RGB to grayscale is a prerequisite to binarization, which ultimately leads to better discrimination of the foreground from the background (i.e., the end goal of text/object detection).   

\subsubsection{Binarization} Binarization methods are conceived to transform grayscale images to binary format (i.e., black and white). The association of each pixel to black or white pixel sets can typically follow a global or a local approach to adaptive thresholding \cite{trier1995goal}. In the former case, the cutoff is constant within the same image, whereas in the latter case the threshold value can vary based on the local illumination conditions of the image. 
Since we could assume a uniform illumination relative to each screenshot, we applied a combination of simple inverse thresholding with Otsu's global binarization.
Finally, we skipped the skew estimation step, given the predominantly horizontal layout of the target text, thus leading to a more scalable pre-processing of the incoming images.
\subsubsection{Segmentation} This step identified rectangular bounding boxes wrapping the textual content of the images, i.e., the standard shape format used by existing OCR engines for text detection \cite{lu1995machine}. 
We adopted a Connected Component based approach (similar to the methodology followed in \cite{talukder2014connected}): (i) the white pixels were first dilated to create more organic white regions (Figure \ref{fig:framework}), (ii) the uniform regions were then detected, and (iii) a rectangle was drawn around the identified area. 
To limit the duplicated recognition of the same regions (i.e., when a smaller rectangle is completely enclosed in a larger area), we included an additional check to filter out the innermost rectangles. However, overlapping rectangles were still identified (Figure \ref{fig:boxover}), leading to partially duplicated text, when bounding boxes were ultimately passed to the OCR engine.

\subsection{Optical Character Recognition (OCR)}
After pre-processing, each segmented region was fed to the OCR engine, using the Python wrapper for Tesseract. 
Tesseract recognizes text in a "two-pass process" \cite{smith2007overview} that integrates character segmentation with the recognition module, and uses backtracking to improve the quality of the output. First, attempts are made to recognize single words separately. Second, part of the words (based on a quality evaluation) are then passed to an adaptive classifier as training data. This increases the ability of the classifier to recognize the remainder of the text in the page. \\
We relied on the stable release of Tesseract (v. 3.03) for our first OCR run, but the alpha release of Tesseract 4.0 was also tested on the considered sample set. The timing of our analysis provided an opportunity to compare the baseline engine with an updated approach, which integrates a LSTM-based module for line recognition in the pipeline, in a similar fashion to the OCRopus framework \cite{breuel2008ocropus}. We were then able to assess the improvement introduced by a Neural-Net component, without increasing the computation and time expenses. Tesseract 4 has already been trained on approximately 400,000 lines of text that include 4,500 fonts. We exploited the pre-trained OCR architecture as-is, without re-training.\\
For each screenshot, we obtain a Unicode text file formatted for easy comparison with the ground truth data.
\subsection{Ground Truth Data Collection}
Gold standard text was needed to ultimately evaluate the quality of the text extracted from the screenshots. Gold standard data is often collected through crowdsourced transcription services (e.g., Amazon Mechanical Turk), or, in some cases, through third-party commercial APIs for text extraction (e.g., Microsoft Cogntive Services). However, as our data are highly sensitive and require privacy protection, we customized a free, and open-source \textit{localturk} tool \cite{dankv} and involved three human annotators trained in human subjects research and confidentiality.

\begin{figure}[t]
\caption{\label{fig:turk}GUI of manual transcription tool (illustrative screenshots).}
\includegraphics[width=\linewidth]{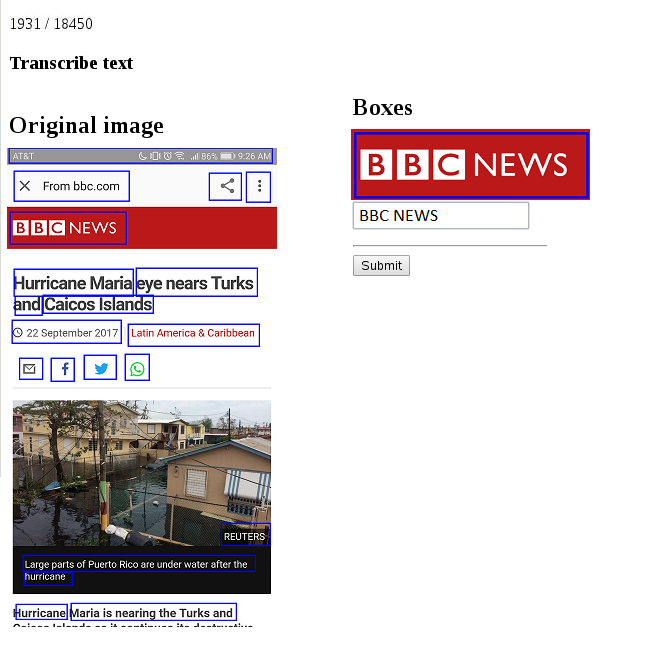}
\end{figure}

The human annotators securely accessed a dedicated Virtual Machine hosting the transcription tool. Figure \ref{fig:turk} showcases the GUI interface used for transcribing the images. The GUI's left side showed the full screenshot, annotated with a set of bounding boxes produced by our Image Pre-processing module, while the individual boxes requiring transcription were displayed on the right side. 
Bounding boxes were presented to annotators in the same scanning order followed by our Image Pre-processing module when detecting the rectangular segments to be passed to the OCR engine, i.e., from top to bottom and from left to right. This precaution ensured consistency between the collected ground truth text and the generated text to be evaluated.\\
The annotators followed detailed instructions to complete the transcription tasks. Specifically, annotators were instructed to preserve capitalization and punctuation, use special character marks to disambiguate icons and graphic contents from text, and notate locations where text spanned beyond the segmentation cuts. The complete image with bounding boxes allowed annotators to check and note if any portions of text had been missed by the bounding box procedure and were not included in the transcription request. Similarly, partially overlapping boxes could arise in the set (Figure \ref{fig:boxover}). Annotators were instructed to take the full picture as reference, to transcribe the overlapping text only once, and to mark any cases that were not covered by the instructions. These checks (i.e., ordering of bounding boxes, handling of overlapping bounding boxes, updating transcription rules dynamically, unanticipated situations) facilitated the subsequent analysis of segmentation inaccuracies.
In this first experimental setup, each annotator transcribed an independent set of images. In other words, agreement across different annotated sets was not evaluated. However, after a first run of evaluation and error analysis, the ground truth collection was cross-checked and manually corrected. This validation step supported the later assessment of the degree of human-error embedded in the process.\\  
Leveraging quality with the expensiveness of the transcription task \cite{SchoningKCAP15}, we obtained complete transcriptions of 1,360 screenshots. In sum, comprehensive ground truth data to evaluate our text extraction pipeline was produced for 10 percent of the analysis sample.

\subsection{Extracted Text Evaluation}
To evaluate the quality of our generated text against the gold standard transcribed text, we defined OCR accuracy both at the character and at the word level, as a complement of the error rates, and further discriminating between order-dependent and order-independent error rate at the word-level evaluation. \\

\textbf{Word Error Rate (WER)} is based on the Levehnstein distance \cite{1966Lev} between the provided text and the reference text:
\begin{equation}
WER=\frac{i_w + s_w + d_w}{n_w}
\end{equation}
where $i_w$, $s_w$ and $d_w$ refer to the number of words to be inserted, substituted and deleted to transform the given text into the reference text. Among all the possible permutations, the values that minimize the sum $i_w$ + $s_w$ + $d_w$ are actually chosen. The resulting number of transformations is then normalized by the number of words in the reference text ($n_w$).\\

\textbf{Character Error Rate (CER)}.
Similarly, the same metric can be defined at the single character level:
\begin{equation}
CER=\frac{i + s + d}{n}
\end{equation}
where word counts are replaced by character counts, following the same rationale defined for the WER metric. Thus, the error rate will typically be higher at the word level, as failures in recognizing single characters ultimately impact the recognition of words as a whole.\\

\textbf{Position-independent WER (PER)} is based on a bag-of-words approach that does not take the provided order of words into account when evaluating the word-level error rate. Hence, this metric loosens one of the constraints of the standard WER. As a result, this metric tends to overestimate the actual word-level accuracy.

Hence, improving the accuracy of our text extraction framework essentially translates into minimizing the error rates, both at the character and at the word level. Furthermore, we prioritize the improvement on the standard WER over its positional-independent counterpart, with the intent to preserve the syntactic and semantic attributes of the extracted text, and support meaningful Natural Language Processing and Topic Evolution analyses of the obtained documents.\\
To record these three metrics, we used the ocrevalUAtion open source tool \cite{carrasco2014open}. Besides the described metrics, this tool also reports error rates by character, and aligned bitext for each document match, facilitating the comparison of the generated output against the reference text \cite{carrasco2014open}. We set up the evaluation so that punctuation and case differences were counted as well, when calculating the scores.\\
Finally, computational time was recorded and evaluated as an additional factor contributing to the overall process efficiency. 

\subsection{Image Retrieval System}
\subsubsection{Document Parsing}
To provide an organized structure to the data set and to link the extracted contents with each source image file, we mapped the image features and metadata into a defined XML schema, consisting of: (i) a unique identifier (i.e., obtained by concatenating each subject ID with the date and time of screen capture), (ii) a timestamp marking the date and time when the screenshot was taken, (iii) the text extracted from the images through OCR (Section 3.4), when present, (iv) an optional category describing the activity depicted by the considered frame (i.e., whenever the considered image had been manually classified), and (v) two file paths linking to previous and next image on the same subject's timeline. The parsing module was developed in Python, exploiting the built-in ElementTree library. XML documents we formatted in a compliant format with Apache Solr, the open source software used to index our collection. 
An example of resulting data structure is as follows:
\begin{verbatim}
<add>
    <doc>
    <field name="id"></field>
    <field name="timestamp"></field>
    <field name="category"></field>
    <field name="text"></field>
    <field name="previous_image"></field>
    <field name="next_image"></field>
    </doc>
</add>
\end{verbatim}
\subsubsection{Indexing}
The XML documents produced in the previous step were then indexed through the Apache Solr 6.3 distribution, which follows an inverse indexing approach. As opposed to a forward index, an inverted index consists of keyword-centric entries, referencing the document containing each considered term. 
While all data fields listed in Section 3.6.1 were stored within the Solr engine, only the textual content embedded in the images and their attached categorical labels were indexed to enable textual search over those attributes.

\subsubsection{Similarity Scoring and Ranking}
Once the index is built, user queries are compared against the collection based on a pre-determined similarity definition, ultimately impacting the returned results. In this application, we computed the similarity between the input textual queries and the inverted index entries through the Okapi BM25 term-weighting scheme \cite{robertson1995okapi}. For a document $D$ and query $Q$ containing $q_i$,...,$q_n$ keywords, the score is then defined as:
\begin{equation}
\small 
score(D,Q)=\sum\limits_{i=1}^n idf(q_i)\cdot \frac{f(q_i,D) \cdot (k_1 +1)}{f(q_i,D) + k_1 \cdot (1-b+b \cdot \frac{|D|}{avdl})}
\end{equation}
where $avdl$ is the average document length and $k_1$ and $b$ are free parameters (kept to default built-in values in this case). Equation 3 is based on the same underlying principles of the term frequency- inverse document frequency (\textit{tf-idf}) heuristic, when determining the relevancy of words within each considered document. Particularly, the $idf$ score recipient, for each keyword $q_i$ is computed as: 
\begin{equation}
idf(q_i)=\log \frac{N - n(q_i) + 0.5}{n(q_i) + 0.5}
\end{equation}
where $n(q_i)$ represents the number of documents containing q$_i$ and $N$ is the total size of the corpus. 
\textit{Tf-idf} increases proportionally with the frequency of occurrence of a term within a document and is normalized based on the term occurrence throughout the whole corpus. 
However, BM25 has been proven more effective than \textit{tf-idf} in mirroring the user's perceived relevancy towards returned contents, due to the introduced regularization parameters \cite{robertson1995okapi}. \\
Further, the aforementioned approach was combined with Solr built-in Query Boosting function, to purposefully increase the ranking of term-matching on categorical fields over term-matching on the screenshot embedded contents. For instance, if the user provided the keyword \textit{"Web"}, multiple candidate documents might be selected, solely based on the BM25 scores. However, if one document was labeled with the term \textit{"Web"}, that document would obtain a higher ranking in the returned result list. 

\subsubsection{Search engine GUI}
\begin{figure}[h]
\caption{\label{fig:searchgui}Search Engine Graphic User Interface(illustrative screenshots).}
\includegraphics[width=\linewidth]{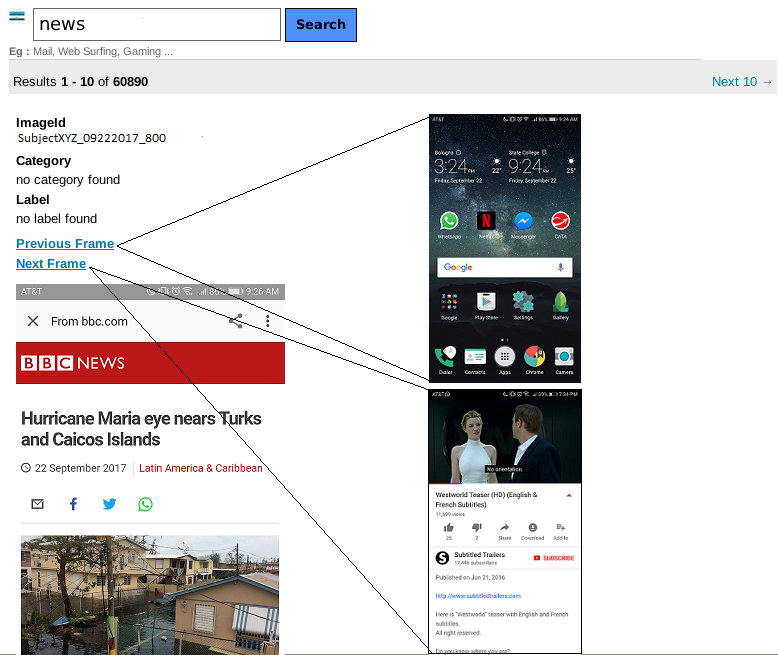}
\end{figure}

The results were presented to the users through a dedicated interface, integrating the Solr-based back-end with the Python-based Django Web framework. Snippets of the developed Graphic User Interface are shown in Figure \ref{fig:searchgui}. For each retrieved image, certain metadata are listed, including the id, timestamp and category. Each thumbnail can be expanded to full-size resolution, aiding further exploration of the retrieved textual content. Links to the previous and following frame on the same subject's timeline emphasize the potential for temporal zooming and longitudinal analyses offered by this novel data set \cite{reeves2017screenomics}, which provides domain-independent elements of sequence, context and interdependence.  


\section{OCR Results and discussion}

As introduced in Section 3.4, we compared the performance of two different releases of the Tesseract OCR engine before and after validating the manually transcribed text. This experimental setup aimed to quantify the improvement introduced by a NeuralNet-based module for line recognition (i.e., the additional component introduced with Tesseract 4 compared to Tesseract 3.03) when the other parameters were kept constant. 

\begin{figure*}[t!]
    \centering
    \begin{subfigure}[t]{0.5\textwidth}
        \centering
        \includegraphics[scale=0.3]{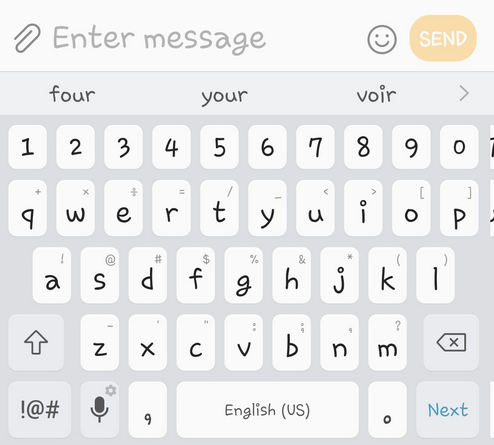}
        \caption{\label{fig:ccooky}}
    \end{subfigure}%
     ~
    \begin{subfigure}[t]{0.5\textwidth}
        \centering
        \includegraphics[scale=0.8]{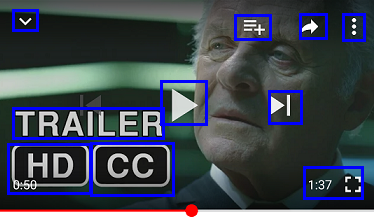}
        \caption{\label{fig:videof}}
    \end{subfigure}
    \centering
    \begin{subfigure}[t]{0.5\textwidth}
        \centering
        \includegraphics[scale=0.3]{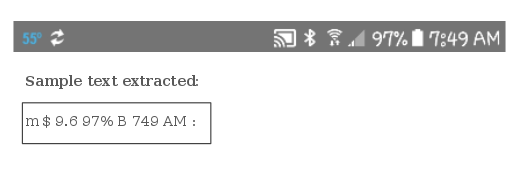}
        \caption{\label{fig:banner}}    
        \end{subfigure}%
     ~   
     \begin{subfigure}[t]{0.5\textwidth}
        \centering
        \includegraphics[scale=0.8]{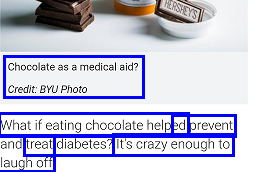}
        \caption{\label{fig:boxover}}
    
    \end{subfigure}
    \caption{Examples of discovered patterns after error analysis (illustrative screenshots): (a) fancy fonts, (b) text embedded in videos is extracted more effectively when integrating the pre-processing routine, (c) smartphone upper banners add marginal noise, (d) inaccurate segmentation can lead to overlapping bounding boxes.}
\end{figure*}

First, we compared the two Tesseract-based solutions, applied as-is, with our framework, which added the aforementioned Image Pre-processing steps upfront. As illustrated in Tables \ref{result1} and \ref{result2}, the Image Pre-processing framework proposed here improved the overall accuracy, both at the character and word level \cite{chiatti2017kcap}. Hence, this improvement justifies the additional computational steps implied by the proposed solution.\\
After a careful inspection and error analysis of the outputs produced in the two cases, we identified a lower robustness of the pre-processing framework in the presence of peculiar fonts chosen as a default on users' phones (an example of such cases is shown in Figure \ref{fig:ccooky} ), as opposed to the baseline (i.e., non pre-processed) alternatives. This observation suggested that the binarization and segmentation parameters can be further fine-tuned to improve the handling and recognition of font sets that are used by particular individuals.
On the other hand, the proposed framework enhanced the recognition of text that is embedded in video frames (as depicted in Figure \ref{fig:videof}), when compared to the baseline performance of the two Tesseract releases. 

\begin{table*}[t]
\caption{\label{result1}Comparison of baseline Tesseract 3 and 4: before and after applying the noise removal heuristic.}
\centering
\begin{tabular}{*7l}
\toprule
\textbf{Approach} & \multicolumn{2}{|c|}{\textbf{Character-level}} & \multicolumn{4}{c}{\textbf{Word-level}} \\
{} & \multicolumn{1}{|l}{ER} & \multicolumn{1}{l|}{Accuracy} & ER & Accuracy & PER & Accuracy \\
\hline
\midrule
\multicolumn{1}{l|}{Baseline Tesseract 3} &  36.13\% & \multicolumn{1}{l|}{63.87\%} & 47.21\% & 52.79\% & 38.78\% & 61.22\%\\
\multicolumn{1}{l|}{Baseline Tesseract 3 + heuristic} &  33.45\% & \multicolumn{1}{l|}{66.55\%} & 43.42\% & 56.58\% & 36.01\% & 63.99\%\\
\multicolumn{1}{l|}{\textbf{Baseline Tesseract 4}} & \textbf{33.68\%} & \multicolumn{1}{l|}{\textbf{66.32\%}} & 
\textbf{41.54\%} & \textbf{58.46\%} & \textbf{33.93\%}& \textbf{66.07\%}\\
\multicolumn{1}{l|}{\textbf{Baseline Tesseract 4 + heuristic}}&  \textbf{31.73\%} & \multicolumn{1}{l|}{\textbf{68.27\%}} & \textbf{38.38\%} & \textbf{61.62\%} & \textbf{31.46\%} & \textbf{68.54\%}\\
\bottomrule
\end{tabular}
\end{table*}

\begin{table*}[t]
\caption{\label{result2}Comparison of Tesseract 3 and 4 with Image Pre-processing: before and after applying the noise removal heuristic.}
\centering
\begin{tabular}{*7l}
\toprule
\textbf{Approach} & \multicolumn{2}{|c|}{\textbf{Character-level}} & \multicolumn{4}{c}{\textbf{Word-level}} \\
{} & \multicolumn{1}{|l}{ER} & \multicolumn{1}{l|}{Accuracy} & ER & Accuracy & PER & Accuracy \\
\hline
\midrule
\multicolumn{1}{l|}{ImgProc+Tesseract 3} &  33.02\% & \multicolumn{1}{l|}{66.98\%} & 44.50\% & 55.50\% & 39.63\% & 60.37\%\\
\multicolumn{1}{l|}{ImgProc+Tesseract 3 + heuristic} &  31.95\% & \multicolumn{1}{l|}{68.05\%} & 41.26\% & 58.74\% & 37.33\% &  62.67\%\\
\multicolumn{1}{l|}{\textbf{ImgProc+Tesseract 4}} & \textbf{31.15\%} & \multicolumn{1}{l|}{\textbf{68.85\%}} & 
\textbf{40.27\%} & \textbf{59.73\%} & \textbf{35.56\%}& \textbf{64.44\%}\\
\multicolumn{1}{l|}{\textbf{ImgProc+Tesseract 4 + heuristic}}&  \textbf{30.68\%} & \multicolumn{1}{l|}{\textbf{69.32\%}} & \textbf{38.95\%} & \textbf{61.05\%} & \textbf{35.06\%} & \textbf{64.94\%}\\
\bottomrule
\end{tabular}
\end{table*}

\begin{table*}[h] 
\caption{\label{result3}Comparison of Tesseract 3 and 4 with Image Pre-processing, after correcting the human-annotated scripts}
\centering
\begin{tabular}{*7l}
\toprule
\textbf{Approach} & \multicolumn{2}{|c|}{\textbf{Character-level}} & \multicolumn{4}{c}{\textbf{Word-level}} \\
{} & \multicolumn{1}{|l}{ER} & \multicolumn{1}{l|}{Accuracy} & ER & Accuracy & PER & Accuracy \\
\hline
\midrule
\multicolumn{1}{l|}{ImgProc+Tesseract 3} &  27.42\% & \multicolumn{1}{l|}{72.58\%} & 39.12\% & 60.88\% & 33.81\% & 66.19\%\\
\multicolumn{1}{l|}{ImgProc+Tesseract 3 + heuristic} &  27.35\% & \multicolumn{1}{l|}{72.65\%} & 37.67\% & 62.33\% & 32.09\% &  67.91\%\\
\multicolumn{1}{l|}{\textbf{ImgProc+Tesseract 4}} & \textbf{25.16\%} & \multicolumn{1}{l|}{\textbf{74.84\%}} & 
\textbf{33.85\%} & \textbf{66.15\%} & \textbf{28.77\%}& \textbf{71.23\%}\\
\multicolumn{1}{l|}{\textbf{ImgProc+Tesseract 4 + heuristic}}&  \textbf{25.69\%} & \multicolumn{1}{l|}{\textbf{74.31\%}} & \textbf{34.38\%} & \textbf{65.62\%} & \textbf{28.76\%} & \textbf{71.24\%}\\
\bottomrule
\end{tabular}
\end{table*}

Further, we wanted to discriminate between the accuracy deficiencies caused by the human error inherent to the transcription process and the error rates directly associated with the adopted Image Processing framework.
Table \ref{result2} shows the results obtained when applying Tesseract 3.03 and Tesseract 4 to the pre-processed smartphone screenshots, i.e., after segmenting the regions of interest that were candidates for carrying fragments of textual content. The introduction of the LSTM-based line-recognition component slightly improved the OCR accuracy both at the single-character and word level.

Analysis of the returned errors showed that the most prominent faults seem associated with: (i) the presence of \textit{icons} and \textit{other graphic features} in line with the text, (ii) defects in the \textit{reference transcriptions}, (iii) presence of \textit{peculiar fonts}, (iv) textual contents that are \textit{difficult to distinguish} from their backgrounds (e.g., both are a light color), and (v) \textit{partially overlapping segmented regions} leading to duplicated characters (as described in Section 3.3.3 - Figure \ref{fig:boxover}). 

To quantify the incidence of the first category of errors, we developed a naive heuristic to post-process our text and filter out the first line when it matches specific regular expressions that plausibly represent the top banner of smartphone displays. 
An example of the critical regions and text generated by the OCR engine when text was mixed with icons is provided by Figure \ref{fig:banner}.
Specifically, top lines were removed only when other textual content besides the upper banner was found. This process elicited a better measure of net accuracy on the textual contents of interest, by eliminating marginal and noisy content. However, when the top banner (i.e., typically icons, clock and battery level) was the only textual information included in the screenshot, it was not filtered out. Overall, applying this heuristic provided a more reliable proxy of the actual accuracy obtained by the current framework.

Inherent human error associated with our manual transcription procedures also contributed to quality loss. Thus, the produced transcriptions were manually validated and the evaluation step repeated after correction. The results, depicted in Table \ref{result3}, demonstrate the significant incidence of inadequate reference text on the overall scores, when compared to Table \ref{result2}. Typical transcription faults include the occurrence of typos or oversights leading to a partial transcription of text that was actually present on the image. As a result of partial transcriptions, text, which was correctly recognized by the OCR engine, was absent from the reference text, artificially increasing the error rate. These inaccuracies were corrected through a posterior validation check. Please note, part of the error and burden related to transcriptions, was caused by the transcription tool's GUI and procedural instructions, which need further refinement, based on the observations collected during this exploratory phase.

After removing the transcription error effect from the set, the solution which integrates an LSTM-based line-recognition system still provided the highest performance. Tests with Tesseract 3.03 and 4 were run in parallel on two identical Debian, quad-core Virtual Machines. \\
The computation times to process the sample (i.e., 13,172 phone screenshots), without applying \textit{ad hoc} training in either of the two cases, were comparable for all pipelines. In sum, there were not any notable tradeoffs between Tesseract 3.03 and 4.0 in terms of process efficiency. 


\section{CONCLUSIONS AND FUTURE WORK}

This paper introduced a complete workflow for text extraction and retrieval from smartphone screenshots. The pipeline is based on OpenCV image-processing and Tesseract OCR modules. We evaluated the quality of the extracted text, and showed how word and character accuracy improved through refinement of image pre-processing procedures and NeuralNet based line-recognition system introduced in the newly released Tesseract 4.0. Detailed analysis of word and character errors suggest that further improvements are possible, both generally and in the data production process. Additional error analyses identified and isolated the most prominent factors contributing to quality loss. Ultimately, a search engine application was developed based on the inherent characteristics of the data at hand, for the immediate use for the involved analysts. 

Human error embedded in the ground truth data production process was present and unanticipated. The findings that correction of human errors provided improvements in accuracy that were of similar size as other technical refinements suggests some reconsideration of how ground truth (and training) data are produced for new data streams. Given the costs, in terms of time and accuracy, there is much incentive to develop iterative solutions that reduce human involvement in the loop to  correction of automatically-generated transcriptions (following a similar approach to the one described in \cite{SchoningKCAP15} for video annotations). 

Future work on the more technical aspects of the text extraction process for screenshots include fine-tuning and sensitivity analysis of the Image Pre-processing parameters, which should increase the solution's robustness in the presence of diverse fonts, embedded icons, and mixed graphic contents. 
Preprocessing may be particularly important for text extraction from laptop screens, for which the method is generalized, due to the possibility that multiple windows may be visible in a laptop screenshot. Problematically, each window might have different background and foreground contrasts, thus significantly introducing marginal noise. Thus, methods that effectively and accurately partition and segment the main active window will need to be developed and refined. As well, the results of this study suggest there is some need for re-training Tesseract-based solutions so that they better handle characters and words that only appear in specific sub-portions of the data (e.g., individual users' device preferences and idiosyncratic font use). These additions and accommodations will expand computation complexity and time and therefore need to be evaluated with respect to added value.\\
We further intend to make our data and experimental procedures as accessible as possible. Given inherent the privacy concerns, we will evaluate and use Web-scraping technologies to form a publicly-available sample data set for testing the implemented workflow. 

In sum, this paper describes the procedures followed for extracting and indexing text from a new kind of document - digital screenshots - that appears to hold high value for studying human behavior. As the pioneer exploration into extracting useful information from the first screenshot repository, the application of Image Pre-processing, OCR and CBIR tools was successful. While we are only at the beginning of learning what these data hold, these initial results are promising and produce excitement about how Information Extraction methods may be adapted for these data and contribute to new knowledge about how, when, and why individuals engage with digital life. 



\section*{ACKNOWLEDGMENTS}
The authors would like to acknowledge the Center for Advanced Study of Behavioral Science at Stanford University where the project was initially developed and the direct funders of this project: the Stanford University Cyber Social Initiative, the Knight Foundation, and the Pennsylvania State University Colleges of Informations Sciences \& Technology and Health \& Human Development.

\bibliographystyle{IEEEtran}
\bibliography{cyberarXiv}

\end{document}